\documentclass[12pt,a4paper]{article}

\usepackage[utf8]{inputenc}
\usepackage[T1]{fontenc}
\usepackage[margin=2.5cm]{geometry}
\usepackage[numbered]{bookmark}
\usepackage{mathtools}
\usepackage{amssymb}
\usepackage{graphicx}
\usepackage[font=small,labelfont=bf]{caption}
\usepackage{authblk}
\usepackage{cite}
\usepackage{dsfont}
\usepackage{xcolor}
\usepackage[titles]{tocloft}

\usepackage{youngtab}
\usepackage{ytableau}

\usepackage{upgreek}

\numberwithin{equation}{section}
\allowdisplaybreaks

\long\def\del#1\enddel{}

\newcommand{\eq}[1]{\begin{equation} #1 \end{equation}}
\newcommand{\al}[1]{\begin{align} #1 \end{align}}
\newcommand{\ml}[1]{\begin{multline} #1 \end{multline}}

\newcommand{\pa}{\partial}

\newcommand{\ket}[1]{|#1\rangle}
\newcommand{\bra}[1]{\langle#1|}
\newcommand{\braket}[1]{\langle#1 \rangle}

\begin{document}

\title{\vspace{2cm}\textbf{On some (integrable) structures in low-dimensional holography}}
\author[*,$\dagger$]{R.~C.~Rashkov}
\affil[*]{\textit{Department of Physics, Sofia University,}\authorcr\textit{5 J. Bourchier Blvd., 1164 Sofia, Bulgaria}\vspace{5pt} \authorcr{ and} \vspace{5pt}} 
\affil[$\dagger$]{\textit{Institute for Theoretical Physics, Vienna University of Technology,}\authorcr\textit{Wiedner Hauptstr. 8--10, 1040 Vienna, Austria}
	\authorcr\vspace{10pt}\texttt{rash@hep.itp.tuwien.ac.at}
	\vspace{1cm}}
\date{}

\maketitle

\begin{abstract}
	
Recent progress in holographic correspondence uncovered remarkable relations between 
key characteristics of the theories on both sides of duality and certain integrable models.
	
In this note we revisit the problem of the role of certain invariants in low-dimensional holography. As motivating example we consider first the entanglement entropy in 2d CFT and projective invariants. Next we consider higher projective invariants and suggest generalization to higher spin theories. Quadratic in invariants deformations is considered and conjectured to play role in low-dimensional higher spin holography. 
\end{abstract}

\vspace{1cm}
\textsc{Keywords:} holographic correspondence, integrable structures, 2d CFT, higher spins
\newpage

\section{Introduction}

One of recent lines of developments suggest that gravitational theories are holographic. Another point of view is that space-time itself is emergent. Both of these approaches have roots in string/gauge theory duality.
String(gravity)/gauge theory duality, along with its tremendous success, poses also
a number of conceptual issue.
One of them is how the Anti-de-Sitter/Conformal Field Theory (AdS/CFT), or holographic correspondence can be explicitly realized? The holographic correspondence means that the (quantum) gravity should be somehow encoded in the boundary theory! How then this information is stored on the boundary and how can it be extracted? It is worth to note that, along with many other things, one of the most attractive features of this duality is that it offers a way to resolve many puzzles of the weak/strong coupling phenomena.

Obviously the answers of these questions could affect some of the foundations of
the current paradigms. Thus, the understanding of these issues is of crucial importance.
The interpretations on both sides of duality unavoidably involves investigation of entanglement entropy at macro- and microscopic level. Recent progress in holographic correspondence uncovered remarkable relations between entanglement entropy (EE) and a specific extremal surface area. The celebrated holographic
entanglement entropy formula states that an extremal surface anchored to the boundary contains the complete information about entanglement entropy for arbitrary regions \cite{Ryu:2006bv}. It is tempting to consider the space of entanglement entropies however, this spaces is obviously much bigger that the space of asymptotically AdS metrics. Thus, to develop effective method for explicit description of theories on both sides it is better to consider geometries dual to holographic states. This line of considerations is based on the assumption that the bulk content represents, in possibly highly non-trivial way, the evolution of particular boundary data. The
latter basically consists of correlation functions of certain local operators. Thus, studying some non-local properties like entanglement entropy in this context is interesting and challenging task. 

While about holographic duality from bulk to boundary direction there are
a number of impressive results, the opposite direction, namely bulk reconstruction
just recently started to attract good deal of attention.
Indeed, to reconstruct bulk theory one has to have complete dictionary between the two sides of the duality. This however is not enough, since we should know how bulk data is encoded into boundary theory. 
The main playground for this type of investigations is the AdS/CFT correspondence and its modifications.
The most tractable case of holographic duality is in the case of (asymptotically)$AdS_3/CFT_2$ where integrable structures play essential role. An inspiring work in this direction for instance is \cite{Perez:2016vqo}. The advantage is that the theories on the boundary are usually integrable even at quantum level. This allows to make conclusions for the dual theory for wide range of couplings. The entanglement entropy in two dimensions is usually calculated by making use of replica-trick (see for instance \cite{Calabrese:2004eu} and references therein). In view of applications to holography, it is important to collect as much as possible information that could be used for bulk theory reconstruction. Thus, study the entanglement entropy of excited states provides another piece complementing that information. In particular, it was shown that for a single interval the entanglement entropy equals to the one-point function of energy momentum tensor (see, for instance, \cite{Beach:2016ocq}).  Looking for projective invariants we find that entanglement entropy of excited states,
calculated via replica-trick method, has an expansion in terms of so-called Aharonov invariants the lowest of which is the Schwarzian \cite{Rashkov:2016xnf}. Excited states can be though of as obtained by certain conformal map.  Following the strategy of looking for integrable structure it is natural to express this map in terms of Faber polynomials. The later have as coefficients the so called Grunsky coefficients. It is know for some time that Grunsky coefficients are related to integrable hierarchies, namely second derivatives of certain tau function. Exploring this fact, it has been shown in \cite{Rashkov:2016xnf} that  entanglement entropy is expressible  as series with coefficients second derivatives of dispersionless
Toda (dToda) tau-function. We should note that Toda theory appears in many physical models (including holographic and higher spin theories) and we conjecture that entanglement entropy is intimately related to Toda hierarchy. 

The very short discussion above points towards importance of the integrability in holographic correspondence not only as a tool, but also as conceptual element in understanding this phenomenon.  
In this note we revisit the problem of the role of conserved quantities and some other characteristics in low-dimensional holography with the idea to generalize the considerations to higher spin theories.  We will take a different view to the problems, namely we will consider integrable structures underlying the main features appearing in this type holography. 

The notes are organized as follows. In the next section we give a brief overview of the integrable structures in expressions like entanglement entropy etc. with an emphasis to projective structures. The results give a strong motivation for more thorough considerations with the idea for generalizations to higher spin holography.  Next, we introduce higher projective structures and represent them in terms of (partial) Bell polynomials asa well as in terms of certain tau-functions. A brief analysis how these invariants could appear in the higher spin actions is offered. We suggest also quadratic deformations for higher spin theories analogous to the so-called $T\bar{T}$, $T\bar{J}$ etc. deformations. 

In the concluding section we comment on the results and some further developments.


\section{Entanglement entropy and projective/integrable structures}\label{sect-2}

One of the central issues in holographic correspondence is the bulk theory reconstruction from boundary data. It is well known that for this purpose not only the correlation functions are needed but also the conformal blocks. Part of the necessary boundary data comes from the excited states.  

Focusing to 2d CFT, the action of the conformal group includes full Virasoro algebra. The identity operator has as a descendant the stress-tensor $T(z)$ (which is quasi-primary)  as well as its derivatives
\eq{
	\mathrm{Id}\sim 1, T, \pa^k T, T^2, T\pa^l T, \cdots.
}
According to the holographic conjecture these are the states that capture the gravitational sector of the dual theory. The excited states play analogous role in the correspondence.

Concerning bulk reconstruction, Ryu-Takayanagi conjecture has a central role. Consider a spacial subsystem $A$ of the CFT and let $S(A)$ is its entropy. The Ryu-Takayanagi formula states the equality
\eq{
S(A)=\frac{1}{4G_N}\operatorname{Area}(\tilde{A}),
}  
where $\tilde{A}$ is co-dimension 2 extremal surface with the same boundary as $A$. We turn now to the field theory side calculations focusing on the excited states.

To obtain the excited states we use the operator-state correspondence. In other words, we assume that the excited states are realized by a conformal transformation $z\:\mapsto\:w=f(z)$ is realized by an unitary operator $U_f$ ($U_fU_f^\dagger=1$) \cite{Beach:2016ocq}. 

The vacuum entanglement entropy (EE) has been considered in many papers, see for instance \cite{Holzhey:1994we, Calabrese:2004eu}. The M\"obius transformations (or its universal cover $SL(2,\mathbb{C})$ leave the vacuum invariant. Therefore, the excited states should be obtained by transformations which do not belong to $SL(2,\mathbb{C})$. Under such transformations two-point correlation function of the primaries  $\Phi_\pm$  is
\eq{
	\braket{\Phi_+(z_1)\Phi_-(z_2)}=\left(\frac{\pa f(z_1)}{\pa z_1}\right)^{h_n} \left(\frac{\pa f(z_1)}{\pa z_1}\right)^{h_n} \braket{\Phi_+(f(z_1))\Phi_-(f(z_2))}.
	\label{transf-twist}
}
The entanglement entropy can be calculated by making use of the R\'{e}nyi entropy.
On the other hand, calculation of the R\'{e}nyi entropy usually uses twist fields $\Phi_\pm$ of dimensions $(h_n,\bar{h}_n)=c/24(n-1/n,n-1/n)$ and normalization as in \cite{Calabrese:2004eu}
\eq{
	\exp\left((1-n)S^{(n)}\right)=\braket{\Phi_+(z_1)\Phi_-(z_2)}=\frac{1}{(z_1-z_2)^{2h_n}}.
}
Then, the entanglement entropy is obtained from R\'{e}nyi entropy $S^{(n)}$ (below $\delta$ is an UV cut-off) taking the limit $n\to 1$
\eq{
	S_{vac}=\lim\limits_{n\to 1}S^{(n)}=\lim\limits_{n\to 1}\log(z_1-z_2)^{-2h_n}=\frac{c}{12}\log\frac{(z_1-z_2)}{\delta^2}.
	\label{vac-renyi-1}
}
Assuming that the excited states are obtained as $\ket{f}=U_f\ket{0}$ and the transformation properties \eqref{transf-twist}, one can easily find for the R\'{e}nyi entropy the expression
\al{
	\exp\left((1-n)S^{(n)}_{ex}\right)&=\bra{f}\Phi_+(z_1)\Phi_-(z_2)\ket{f} =
	\braket{0|U_f^\dagger \Phi_+(z_1)U_f\:U_f^\dagger\Phi_-(z_2)U_f|0}	 \label{1-point-1}\\
	&=\left(\frac{df}{dz}\right)_{z_1}^{-h_n}\left(\frac{df}{dz}\right)_{z_2}^{-h_n} \left(\frac{d\bar{f}}{d\bar{z}}\right)_{\bar{z}_1}^{-\bar{h}_n}  \left(\frac{d\bar{f}}{d\bar{z}}\right)_{\bar{z}_2}^{-\bar{h}_n}
	\bra{0}\Phi_+(f(z_1))\Phi_-(f(z_2))\ket{0},\label{1-point-2}.
}
Thus, the limit $n\to1$ gives the entanglement entropy for the excited states
\eq{
	S_{ex}=\lim\limits_{n\to 1}S^{(n)}_{ex}=-\frac{c}{12}\log\left|\frac{f'(z_1)f'(z_2)\,\delta^2}
	{(f(z_1)-f(z_2))^{2}} \right|.
	\label{ex-renyi-1}
}
It is well known for long time that expressions of this type satisfy Liouville field equation
\eq{
\delta^2\frac{\pa^2 S_{ex}[f]}{\pa z_1\pa z_2}= \frac{c}{6}e^{-\frac{12}{c}S_{ex}[f]}.
\label{lio-1}
}
This is a strong argument to proceed in searching for integrable structures in such important characteristics of the theories on both sides of the duality. Indeed, the first step naturally appears to be looking for $SL(2,\mathbb{C})$ invariants in entanglement entropy of excited states. Expanding the argument of the logarithm in \eqref{ex-renyi-1} in the interval length, the expression actually becomes expansion into the so-called Aharonov invariants \cite{Rashkov:2016xnf}-\cite{harmelin}
\al{
\frac{f'(z)f'(\zeta)}{(f(\zeta)-f(z))^2} &=\frac{1}{(z-w)^2}+\frac{1}{6}S(f)(z)+\frac{1}{12}S'(f)(z)(z-w)+\cdots \nonumber \\
& =\frac{1}{(\zeta-z)^2}+ \sum\limits_{n=2}^\infty(n-1)\psi_n[f](z)(\zeta-z)^{n-2}.
\label{aharon-3}
}
The expression is invariant under M\"{o}bius transformations $M$, $M\circ f(z)$
and thus, 
\eq{
\psi_n[M\circ f]=\psi_n[f], \qquad n\geq 2.
}
The quantities $\psi_n[f]$ are called Aharonov invariants and satisfy the recursion formula \cite{aharonov} (see also \cite{meira,harmelin}):
\eq{
(n+1)\psi_n[f]=\psi_{n-1}[f]'+\sum\limits_{k=2}^{n-2}\psi_k[f]\psi_{n-k}[f], \quad n\geq 3.
\label{aharon-5}
}
The first several invariants are
\eq{
\psi_1[f]=\frac{1}{2}\frac{f''(z)}{f'}, \quad \psi_2[f]=\frac{1}{3!}S(f), \quad \psi_3[f]=\frac{S(f)'}{4!}; \quad \psi_4=\frac{S''(f)}{5!}+\frac{S^2(f)}{5(3!)^2},
\label{aharon-few}
}
where $S(f)$ is the Schwarzian derivative
\eq{
S(f)\equiv \{f,z\}=\frac{f'''}{f'}-\frac{3}{2}\left(\frac{f''}{f'}\right)^2.
\label{schwarzian-a}
}
In \cite{meira} it was proven that all M\"obius invariants are derivable from the Schwarzian derivative $S(f)$.

The next natural step is to explore the fact that the expression for entanglement entropy satisfies the Liouville equation. The later is the first member of the Toda fields family and  it would be natural to expect integrable structures of this hierarchies to appear in our expression \eqref{ex-renyi-1} To find a relation between this expression and Toda hierarchy in is instructive to expand on the interval endpoints. Before doing that, we should stress on that the confomal maps must be univalent functions, one univalent around the origin (infinite past) and the other in neighborhood of infinity (infinite future). These have expansions as follows
\al{
	& \tilde{S}=\left\{f(z)=a_1z+a_2z^2+a_3z^3+\cdots=\sum\limits_{n=1}^\infty a_nz^n,\: a_1\neq 0 \right\}
	\label{faber-16a} \\
	& \Sigma=\left\{g(z)=z+b_0+\frac{b_1}{z}+\cdots=bz+\sum\limits_{n=0}^\infty b_nz^{-n} \right\}.
	\label{faber-16b}
}
Let $f\in\tilde{S}$ is univalent in a neighborhood $V$ of $0$ and $g\in\Sigma$ is univalent in a neighborhood $U$ of $\infty$. It is said that the pair $(f,g)$ are disjoint relative to $(V,U)$ if the sets $f(V)$ and $g(U)$ are disjoint. 
Thus, the functions
\eq{ \log\frac{g(z)-g(\zeta)}{z-\zeta}, \qquad \log\frac{g(z)-f(\zeta)}{z-\zeta}, \qquad \log\frac{f(z)-f(\zeta)}{z-\zeta},
	\label{univalent-0}
}
are analytic in $U\times U$, $U\times V$ and $V\times V$ respectively. For our purpose we need to expand  $g'(z)/(g(z)-w))$ in series with respect to $z$ at $\infty$
\eq{
	\frac{g'(z)}{g(z)-w}=\sum\limits_{n=0}\Phi_n(w) z^{-n-1}, \qquad \Phi_0(w)\equiv 1.
	\label{faber-2}
}
The polynomials $\Phi_n(w)$
\eq{
	\Phi_n(w)=\sum\limits_{m=0}^nb_{n,m}w^m,
}	
are called Faber polynomials and $b_{nm}$ are called Grunsky coefficients. They satisfy the recursion relation
\eq{
	\Phi_{n+1}(w)=(w-b_0)\Phi_n(w)-\sum\limits_{k=1}^{n-1}b_{n-k}\Phi_k(w)+(n+1)b_n.
	\label{faber-5}
}
It is a simple exercise to find differential relations between the Faber polynomials, which we present below for completeness
\eq{
	\frac{1}{k}\Phi_k''(w)=\sum\limits_{n+m=k}\frac{1}{nm}\,\Phi_n'(w)\Phi_m'(w).
	\label{faber-der-3}
}
The coefficients $b_{nm}$ are called Grunsky coefficients. They are symmetric in its indices and polynomials in $b_k$’s. 
Using above properties of the univalent functions, one can write the non-singular part as an expansion with Grunsky coefficients \cite{Rashkov:2016xnf}. Indeed, having defined Grunsky coefficients, one can write \eqref{aharon-3} as \cite{Wiegmann:1999fr}
\eq{\label{kernel-2}
	\frac{f'(z)f'(w)}{(f(z)-f(w))^{2}}\,-\,\frac{1}{(z-w)^{2}}=\frac{\partial ^{2}}{\partial z\partial w}\log \frac{f(z)-f(w)}{z-w}=-\sum _{m,n\geq 1}mn\,b_{mn}z^{-m-1}w^{-n-1}.
}
Therefore, the difference between the vacuum entanglement entropy and that of excited states\eqref {ex-contrib-1} can be written as\footnote{As a side remark, the direct calculation \eqref{aharon-3} allows to find the Schwarzian in terms of Grunsky coefficients 
\eq{
\label{kernel-4}
\frac{1}{6}S(f)(z)=-\frac{1}{z^2}\sum _{m,n\geq 1}mn\,b_{mn}z^{-n-m}.	
}}
\eq{
	S_{vac}-S_{ex}=\frac{c}{12}\log\left(1+(z-w)^2
	\sum _{m,n\geq 1}mn\,b_{mn}z^{-m-1}w^{-n-1}
	\right).
	\label{schwarzian-grunsky-fin}
}
Now we are ready to make a link to Toda dispersionless hierarchy.

To describe Toda dispersionless hierarchy we will use the 
Sato approach to integrable hierarchies, namely the formalism of pseudodifferential operators, $\mathcal{L}=\pa+ \sum_{i=1}^{\infty}u_i\pa^{-i+1}$. Then
the dispersionless Toda hierarchy is defined as the system of differential equations
\al{
	& \frac{\pa\mathcal{L}}{\pa t_n}=\{\mathcal{B}_n,\mathcal{L}\}, \quad
	\frac{\pa\mathcal{L}}{\pa t_{-n}}=\{\mathcal{\bar{B}}_n,\mathcal{L}\}, \label{dtoda-zakh-1}\\
	& \frac{\pa\mathcal{\bar{L}}}{\pa t_n}=\{\mathcal{B}_n,\mathcal{\bar{L}}\}, \quad
	\frac{\pa\mathcal{\bar{L}}}{\pa t_{-n}}=\{\mathcal{\bar{B}}_n,\mathcal{\bar{L}}\}, \label{dtoda-zakh-2}
}
where $\mathcal{L}$ and $\mathcal{\bar{L}}$ are generating functions of unknowns  $u_i = u_i(t;s)$, $\bar{u}_i=\bar{u}_i(t;s)$. To make a link to the previous considerations it is useful instead of using pseudodifferential operators to work in terms of their symbols\footnote{With slight abuse of notations we will use the same letters $\mathcal{L}_n$ and $\mathcal{B}_n$ as before.}
\eq{		\mathcal{L}=p+u_1+u_2p^{-1}+u_3p^{-2}+\cdots, \qquad  
	\mathcal{\bar{L}}=\bar{u}_0p^{-1}+\bar{u}_1+\bar{u}_2p+\bar{u}_3p^2+\cdots 
	\label{dtoda-4}
}
The operators $\mathcal{B}_n$, $\mathcal{\bar{B}}_n$ are defined by
\eq{ 
	\mathcal{B}_n = \left(\mathcal{L}^n \right)_{\geq0}, \qquad
	\mathcal{\bar{B}}_n = \left(\mathcal{\bar{L}}^{-n}\right)_{\le 0}.
}
where the truncation operations $()_{\geq 0}$ and $()_{\le 0}$ denote the
polynomial part and the negative degree part in $p$. 
The Poisson brackets for symbols are
\eq{
\{f,g\}=p\frac{\pa f}{\pa p}\frac{\pa g}{\pa t_0}-p\frac{\pa g}{\pa p}\frac{\pa f}{\pa t_0}.
}
A crucial point in the theory is the existence of the so-called tau function, which is related to the free energy of dispersionless Toda system as $\mathcal{F}=\log\tau_{dToda}$. To find explicit expressions we first note that from \eqref{dtoda-4}  one finds
\eq{ 
	p^m = \mathcal{L}^m-u_{m,0} \mathcal{L}^{m-1}-(u_{m,1}-u_{m-1,1}u_{m,0})\mathcal{L}^{m-2}-\cdots.
}
Using standard techniques as in \cite{Teo:2003aw,Wiegmann:1999fr}, one can prove the relations
\al{
	\log p& =\log\mathcal{L}-\sum\limits_{m=1}^\infty\frac{1}{m} \frac{\pa^2\mathcal{F}}{\pa t_0\pa t_m}\mathcal{L}^{-m} \label{toda-tau-2} \\
	\left(\mathcal{L}^n \right)_{\geq 0}& =\mathcal{L}^n- \sum\limits_{m=1}^\infty\frac{1}{m} \frac{\pa^2\mathcal{F}}{\pa t_n\pa t_m}\mathcal{L}^{-m} =
	\frac{\pa^2\mathcal{F}}{\pa t_0\pa t_n}-\sum\limits_{m=1}^\infty\frac{1}{m} \frac{\pa^2\mathcal{F}}{\pa t_{-m}\pa t_n}\mathcal{\bar{L}}^m \label{toda-tau-3} \\
	\log p& =\log\mathcal{\bar{L}}+\frac{\pa^2\mathcal{F}}{\pa t_0^2}-\sum\limits_{m=1}^\infty\frac{1}{m} \frac{\pa^2\mathcal{F}}{\pa t_{-m}\pa t_0}\mathcal{\bar{L}}^{m} \label{toda-tau-4} \\
	\left(\mathcal{\bar{L}}^{-n} \right)_{\le 0}& = - \sum\limits_{m=1}^\infty\frac{1}{m} \frac{\pa^2\mathcal{F}}{\pa t_{-n}\pa t_{m}}\mathcal{L}^{-m}
	=\mathcal{\bar{L}}^{-n}+ \frac{\pa^2\mathcal{F}}{\pa t_0\pa t_{-n}}- \sum\limits_{m=1}^\infty\frac{1}{m} \frac{\pa^2\mathcal{F}}{\pa t_{-n}\pa t_{-m}}\mathcal{\bar{L}}^{m} \label{toda-tau-5}.
}

Now we are in a position to map these quantities to the previous expressions. First we identify $p$ with $w=g(z)$ belonging to the class $\Sigma$ and $\mathcal{L}$ with $z$. Next, $\mathcal{F}(p)$ is identified with $z=G(w)$, i.e. the inverse of $g(z)$. Thus, $\mathcal{B}_n$ become \textit{Faber polynomials}! Reversing the considerations and following for instance \cite{Carroll:1995mu}, one can prove for consistency that the so defined tau function satisfies the Hirota dispersionless equation. The identification we made maps the expression \eqref{kernel-2} to
\eq{\label{kernel-tau}
	\frac{f'(z)f'(w)}{(f(z)-f(w))^{2}}\,-\,\frac{1}{(z-w)^{2}}=\sum _{m,n}
	\frac{\pa^2\mathcal{F}}{\pa t_m\pa t_n}z^{-m-1}w^{-n-1}.
}
The Grunsky coefficients $b_{nm}$ of the pair $(g=w(\mathcal{L}),\: f=w(\mathcal{\bar{L}}))$  are related to the tau function, or free energy as follows:
\al{
	& b_{00}=-\frac{\pa^2\mathcal{F}}{\pa t_0^2}, \quad b_{n,0}= \frac{1}{n}\frac{\pa^2\mathcal{F}}{\pa t_0\pa t_n}, \quad b_{-n,0}= \frac{1}{n}\frac{\pa^2\mathcal{F}}{\pa t_0\pa t_{-n}},\quad n\geq 1 \nonumber \\
	& b_{m,n}= -\frac{1}{mn}\frac{\pa^2\mathcal{F}}{\pa t_m\pa t_n}\qquad b_{-m,-n}= -\frac{1}{mn}\frac{\pa^2\mathcal{F}}{\pa t_{-m}\pa t_{-n}}, \quad n,m\geq 1\label{tau-grunsky} \\
	& b_{-m,n}= b_{n,-m}=-\frac{1}{mn}\frac{\pa^2\mathcal{F}}{\pa t_{-m}\pa t_n},\quad n,m\geq 1. \nonumber
}
In particular, the Schwarzian takes the form
\eq{
	S(f)(z)=\frac{6}{z^2}\sum\limits_{m,n}
	\frac{\pa^2\mathcal{F}}{\pa t_m\pa t_n}z^{-m-n}.
	\label{schwarzian-tau}
}
The higher Aharonov invariants can be easily found in terms of the tau function, or equivalently $\pa^2\mathcal{F}/\pa t_n\pa t_m$. For this purpose it is convenient to introduce the quantity
\eq{
T_j=\sum\limits_{n+m=j}\frac{\pa^2\mathcal{F}}{\pa t_m\pa t_n}\equiv \sum\limits_{n+m=j}\frac{\pa^2}{\pa t_m\pa t_n}\log\tau_{dToda}.
\label{h-tau-1}
}
Thus, the Schwarzian can be written as
\eq{
S(f)(z)=6\sum\limits_{j=1}^\infty T_jz^{j-2}.
\label{h-tau-2}
}
Using the explicit expressions in \eqref{aharon-few} for the first few Aharonov invariants, one finds
\al{
& \psi_2=\sum\limits_{j=1}^\infty T_jz^{j-2}, \quad \psi_3=\sum\limits_{j=1}^\infty (j-2)T_jz^{j-3}, \nonumber \\
& \psi_4= \frac{1}{20}\sum\limits_{j=1}^\infty \left[(j-2)(j-3)T_j +4 Y_j\right]z^{j-4}, \qquad Y_j=\sum\limits_{n+m=j}T_nT_m.
\label{h-tau-3}
}

To conclude this section we note that Liouville and Toda equations appeared in the literature in relation with entanglement entropy, see, for instance comments in \cite{deBoer:2016bov,deBoer:2016pqk}
Thus it is natural to conjecture  that there is deeper sophisticated relations between integrable hierarchies and entanglement entropies in holography.

\section{Higher projective invariants and complex structures}\label{sect-3}

As we mentioned above, we would like to generalize previous considerations to higher spin theories.
In terms of complexity, higher spin gravity is considerably simpler than string theory, since it is basically an ordinary quantum field theory. On the other hand it is much simpler than the full string theory. Having legs on both, gauge theories and String theory, higher spin gravity in 3d provides a very useful laboratory for testing various ideas about weak/strong phenomena.

The generalization we are looking for requires higher projective invariants and we are going to discuss them now.
There are quite some papers on higher projective invariants, including such oriented towards holographic applications to higher spin theories and Sachdev-Ye-Kitaev (SYK) models, see for instance \cite{Li:2015osa,Gonzalez:2018enk,Ecker:2019ocp} and references therein. Here we will specifically focus on the setup allowing to identify the integrable structures underlying the theories as well as to make link to W-algebras.

We start with brief review of the higher projective invariants.  Motivated by the appearance of Schwarzian derivative as a $SL(2,\mathbb{C})$ invariant the holographic structures, we are going to consider the $SL(n,\mathbb{C})$ case.  

The projective invariants are naturally associated with the solutions of the ordinary differential equation
\eq{
	y^{(n)}+ p_{n-2}(z)y^{(n-2)}+\cdots + p_0(z)y=0.	
	\label{ode-w-eq}
}
Suppose we have $n$ linearly independent solutions $(y_1,\,y_2,\dots,\,y_n)$ and define "projective coordinates" in the space of solutions as $f_i=y_i/y_n$ for $i=1,\dots,n-1$. The $\{f_i\}$ can be thought of (locally) as homogeneous coordinates on $\mathbb{CP}^{n-1}$. To construct such invariants we consider a differential equation for the projective "coordinates" $f_i$ which means that the n-th solution is just $1$. This means that in the general case the differential equation contains \textit{only} derivatives but not free term. Therefore,  we should consider the equation 
\eq{
	f^{(n)}+ a_{n-1}(z)f^{(n-1)}+\cdots+a_1(z)f' =0.
	\label{ode-w-12}
}
Again, assuming that $f_i$ are the $n$ linearly independent solutions with $f_n=1$, we find for the Wronskian of this equation
\eq{
	W_n=\begin{vmatrix}f_1' & f_2' & \cdots & f_{n-1}' \\ f_1'' & f_2'' & \cdots & f_{n-1}'' \\ \vdots & \vdots & \ddots & \vdots \\ f_1^{(n-1)} & f_2^{(n-1)} & \cdots & f_{n-1}^{(n-1)}
	\end{vmatrix}.
}

Let us first define 
\eq{
	\tilde{W}_i=\begin{vmatrix}f_1' & f_2' & \cdots & f_{n-1}' \\ \vdots & \vdots & \ddots & \vdots \\
		f_1^{(i-1)} & f_2^{(i-1)} & \cdots & f_n^{(i-1)} \\
		f_1^{(i+1)} & f_2^{(i+1)} & \cdots & f_n^{(i+1)} \\
		\vdots & \vdots & \ddots & \vdots \\
		f_1^{(n)} & f_2^{(n)} & \cdots & f_{n-1}^{(n)}
	\end{vmatrix}, \qquad W_i=(-1)^{n+i}\tilde{W}.
	\label{ode-w-14}
}
The coefficients $a_i(z)$ in \eqref{ode-w-12} are 
\eq{
	a_i=\frac{W_i}{W_n}=(-1)^{i}\frac{\tilde{W}_i}{W_n},
	\label{ode-w-13}
}
Thus, the equation \eqref{ode-w-12} can be written as
\eq{
	f^{(n)} -\frac{W_{n-1}}{W_n}f^{(n-1)}+\cdots+ (-1)^{n+1}\frac{W_1}{W_n}f' =0.
	\label{ode-w-12a}
}
In these notations the projective invariants $q_i$ for $i=0,1,\dots,n-2$ are given by\cite{kim,rashkov-nordita-16} ($p_i\equiv q_i$)
\eq{ q_i=\frac{W_n^{-1}}{\sqrt[n]{W_n}}\left[\sum\limits_{j=0}^{n-i}(-1)^{2n-j}(1-\delta_{nj})\binom{n-j}{n-j-i}
	W_{n-j}\left(\sqrt[n]{W_n}\right)^{(n-j-i)} \right], 
	\label{ode-w-fin}
}
It is known that the coefficients of the equation \eqref{ode-w-12} are invariants, but we would like to construct \textit{projective invariants}!
This means the quantities above are invariant with respect to the transformations of the form
\eq{
	f_i\:\longrightarrow\: \frac{a_{i,0}+a_{i,1}f_1+\cdots +a_{i,n-1}y_{n-1}}{b_0 + b_1f_1+\cdots +b_{n-1}f_{n-1}}, \qquad i=1,2,\dots, n-1.
	\label{ode-w-3a}
}

Although determinant representation is very similar to some structures of the integrable systems, for instance tau-functions, for practical purposes is very useful to give another expression for the invariants. The projective invariants associated to the ordinary differential equation
\eq{
	y^{(n)}+ p_{n-2}(z)y^{(n-2)}+\cdots + p_0(z)y=0,
	\label{ode-w-eq-1}
}
can be written as
\eq{ q_{i>0}=\sum\limits_{j=0}^{n-i}\sum\limits_{k=1}^{n-j-i}(-1)^{2n-j}\frac{(n-j)_i}{\Gamma(i+1)}\left(\frac{1}{n}\right)_k  \frac{W_{n-j}}{W_n^{k+1}} B_{n-j-i,k}\left(W_n',W_n'',\dots,{W_n}^{(n-j-i)}\right) 
	\label{ode-w-fin-2}
}
and 
\eq{ q_{0}=\sum\limits_{j=0}^{n-1}\sum\limits_{k=1}^{n-j}(-1)^{2n-j}\left(\frac{1}{n}\right)_k  \frac{W_{n-j}}{W_n^{k+1}} B_{n-j,k}\left(W_n',W_n'',\dots,{W_n}^{(n-j)}\right) .
	\label{ode-w-fin-2-0}
}
Here $(x)_n$ stands for Pochhammer symbols and $B_{n,k}(x_1,x_2,\dots,x_n)$ are partial or incomplete exponential Bell's polynomials\footnote{Short information about Bell's polynomials and some notations are given in the Appendix.}. Since Bell's polynomials $B_{n,k}$ are homogeneous polynomials of degree $k$, it is convenient to define the quantities
\eq{
\omega_k=\frac{W_k}{W_n}, \qquad 	\hat{w}_n^{(l)}=\frac{W^{(l)}_n}{W_n}, \qquad W_n^{(l)}=\frac{d^lW_n}{dx^l}, \quad k,l=1,\dots.
}
Thus, the above invariants can be written in the form
\eq{ q_{i>0}=\sum\limits_{j=0}^{n-i}\sum\limits_{k=1}^{n-j-i}(-1)^{2n-j}\frac{(n-j)_i}{\Gamma(i+1)}\left(\frac{1}{n}\right)_k  \omega_{n-j} B_{n-j-i,k}\left(\hat{w}_n',\hat{w}_n'',\dots,{\hat{w}_n}^{(n-j-i)}\right) 
	\label{ode-w-fin-2b}
}
and 
\eq{ q_{0}=\sum\limits_{j=0}^{n-1}\sum\limits_{k=1}^{n-j}(-1)^{2n-j}\left(\frac{1}{n}\right)_k  \omega_{n-j} B_{n-j,k}\left(\hat{w}_n',\hat{w}_n'',\dots,{\hat{w}_n}^{(n-j)}\right)  .
	\label{ode-w-fin-2-0b}
}

As an example, let us consider $n=2$, i.e.
\eq{
	y''+p(z)y=0.
}
For the two linearly independent solutions $(y_1,y_2)$ we define the quantity
$f=y_1/y_2$.
According to the considerations above, the only projective invariant is given by
\eq{
	q_0=\frac{1}{2}\{f,z\},
	\label{sl2-schw}
}
where
\eq{
	\{f,z\}=\frac{f'''}{f'}-\frac{3}{2}\left(\frac{f''}{f'}\right)^2,
}
is the Schwarzian derivative. This, of course is pretty well known result. 

For completeness, here we provide the invariants in the case of $n=3$. This case corresponds to the third order equation
\eq{
	y'''+p_1(z)y'+p_0(z)y=0.
	\label{example-n3-1}
}
In this case there are two projective invariants. The formula \eqref{ode-w-fin} gives\cite{rashkov-nordita-16,rashkov-nordita-18}
\eq{
	q_0=\frac{1}{3}\left[\omega_1\omega_2-\omega_2''-\frac{2}{9}\omega_2^3\right], \quad q_1=\omega_1+\omega_2'-\frac{1}{3}\omega_2^2.
	\label{w3-case}
}
where
\eq{
\omega_1=\frac{W_1}{W_3}=\frac{f_1'''f_2''-f_1''f_2'''
	}{f_1'f_2''-f_1''f_2'}, \qquad \hat{w}'_3=\omega_2=\frac{W'_3}{W_3}=\frac{W_2}{W_3}=\frac{f_1'f_2'''-f_1'''f_2'}{f_1'f_2''-f_1''f_2'}
	\label{w3-case-a}
}
Let us compare our formula with the expression found in \cite{Li:2015osa} for instance. The expression for the energy-momentum tensor in that paper can be written as
\eq{
T =\left\{ \pa\left(\frac{\pa^2e}{\pa e} +\frac{\pa^2f}{\pa f}\right)-\frac{1}{3}\left(\frac{\pa^2e}{\pa e}\right)^2 -\frac{1}{3}\left(\frac{\pa^2f}{\pa df}\right)^2 \right\} +\frac{1}{3} \frac{\pa^2e}{\pa e} \frac{\pa^2f}{\pa f}.
	\label{thei-1}
}
The correspondence between \eqref{thei-1} and our expression is realized by the identification
\eq{
f=f_2, \quad e=\frac{\pa f_1}{\pa f_2}.
\label{thei-2}
}
Then, using that $\pa W_3=W_2$ we find that the two expressions coincide. The comparison with the other expression goes analogously, but is much more lengthy and we skip it here.

It is straightforward to obtain higher invariants using \eqref{ode-w-fin}, but the relation to integrable structures is still somehow more subtle\cite{rashkov-prep}.
Represented in this form however, the expressions\eqref{ode-w-fin} suggests that it may have also representation in terms of tau-functions\cite{rashkov-nordita-18,rashkov-prep}. 

To discuss the geometric meaning of the projective invariants we focus on Chern-Simons theory with $A_n$ symmetry algebra. Starting with the simplest case, namely $SL(2,\mathbb{R})$, let us define first the symplectic manifold $(\mathcal{M},g)$ of \textit{all} 2d $SL(2,\mathbb{R})$ connections
\eq{
	A=A_zdz+A_{\bar{z}}d\bar{z}, 
	\label{1-ward-10}
}
with symplectic form descended from Chern-Simons (CS) action
\eq{
	\Omega = \frac{1}{2\pi\hbar}\int_\Sigma \operatorname{tr}\delta A_z\wedge \delta A_{\bar{z}}.
	\label{1-ward-11}
} 

Below we list some of the main features.

First of all, we remind that the generator of infinitesimal transformations (due to constraint coming from the flatness condition $F(A)=[\pa- A, \bar{\pa}-\bar{A}]=0$) is
\eq{
	H_\epsilon (A)= \int_\Sigma \operatorname{tr}\epsilon F(A).
\label{1-ward-12}
}
	
Next natural step is to make a symplectic reduction. Making that  we find for the reduced symplectic form
\eq{
\Omega_{red} = \frac{1}{2\pi\hbar}\int_\Sigma \delta \mu \wedge \delta T.
\label{1-ward-13}
}
	
 Thus, the flatness condition becomes
\eq{
\left(\bar{\pa}-\mu\pa -2\pa\mu  \right)T=-\frac{1}{2}\pa^3\mu. 
\label{1-ward-14}
}
	
On the other hand, Beltrami equation
\eq{
\left(\bar{\pa}-\mu\pa\right)F=0,
\label{1-ward-15}
}
defines diffeomorphisms
\eq{
(z,\bar{z})\: \longrightarrow (F(z,\bar{z}),\bar{F}(z,\bar{z})), \qquad \mu= \frac{\bar{\pa} F}{\pa F}.
\label{1-ward-16}
}
The diffeomorphism transformations brings the measure to the following form
\eq{
|dF(z,\bar{z})|^2=|\pa F(z,\bar{z})|^2\,|dz+\mu d\bar{z}|^2\quad \Longrightarrow \quad ds^2=e^{\phi}|dz+\mu d\bar{z}|^2.
\label{1-ward-17}
}
Here $\mu$ is the complex structure while $T$ is the projective invariant, $T(z)=S(f)$.

Generalization of this picture has been suggested in the beautiful paper \cite{Bilal:1990wn} (see also \cite{gerasimov-levin-Mars}).
The corresponding generalized complex structure is determined by the ratios of the solutions of
\eq{
\left[\bar{\pa}-\sum\limits_{k=0}^{n-1}a^{(n)}_{n-1,n-1-k}\pa^k \right]\psi=0.
\label{2-ward-6}
}

Let us go back to \eqref{1-ward-10} writing it in more explicit form after the symplectic reduction
\eq{
	A_z=\begin{pmatrix}
		0&1& 0 & 0 & \cdots & 0\\ 0 & 0 & 1 & 0 & \cdots & 0 \\ \vdots & \vdots & \ddots & & & \vdots \\ \vdots & \vdots & & & 1 & 0 \\ 0 & 0 & 0 & & 0 & 1 \\ u_n & u_{n-1} & u_{n-2} & \cdots & u_2 & 0 
	\end{pmatrix},
	\label{2-ward-7}
}
and parametrize the antiholomorphic part by
\eq{
	A_{\bar{z}}= \begin{pmatrix}
		a_{n-1,n-1} & \cdots & a_{n-1,1} & a_{n-1,0}\\ \vdots & & \vdots & \vdots \\ a_{1,n-1} & \cdots & a_{1,1} & a_{1,0} \\ a_{0,n-1} & \cdots & a_{0,1} & a_{0,0}
	\end{pmatrix}.
	\label{2-ward-8}
}
The coefficients $a_{k,0}=\mu_{k+1}$ generalize the Beltrami differentials and since the symmetry algebra is $A_n$, the coefficient $a_{0,0}=-\sum_k a_{kk}$.

To this end we define the differential operator
\eq{
	\mathcal{D}_k=\pa^k-\sum\limits_{l=2}^ku_k\pa^{k-l},\qquad k\leq n,
	\label{2-ward-9}
}
as well as, its formal adjoint
\eq{
	\mathcal{D}_k^\dagger=(-1)^k\pa^k\psi -\sum\limits_{l=2}^k(-1)^{k-l}\pa^{k-l}(u_l\psi)
	\label{2-ward-10}
}

Thus, one can determine the elements with $i>j$ recursively by making use of
\eq{
	a_{ij}= \pa a_{i+1,j} + a_{i+1,j+1} + a_{i+1,0}u_{j+1}, \quad 0\leq i\leq n-1; \quad  1\leq j \leq n-1. 
	\label{2-ward-11}
}
We quote the formula from \cite{Bilal:1990wn}
\eq{
	a_{l,l-p}=\sum\limits_{k=p}^l(-1)^{k-p}\left[\binom{l-p}{k-p}-\binom{l-p-2}{k-p-2} \right]\pa^{k-p}\mu_{k+1}+\sum\limits_{k=p+2}^l\mathcal{D}^\dagger_{k-p} \mu_{k+1} 
} 

The main conclusions from the considerations in \cite{Bilal:1990wn} are:
\begin{itemize}
	\item The $SL(n,\mathbb{C})$-generalized complex structures are in one-to-one correspondence with the $\mu_k(z,\bar{z})$, $k=2,\dots,n$.
	\item $\mu_k(z,\bar{z})$ are (higher) Beltrami differentials, i.e. (-1,k-1)-differentials.
	\item The corresponding generalized complex structure is determined by the ratios of the solutions	of \eqref{2-ward-6}.
	
\end{itemize}

 In the next section we will provide more comments on these quantities.

\section{Deformations of HS theories and projective invariants}\label{sect-4}

To proceed, we briefly remind the most direct approach to the higher spin (asymptotically) $AdS_3/CFT_2$ correspondence.

The CS formulation of 3d gravity is the appropriate setup for HS generalizations. While the spin two case is based on $SL(2)$ symmetry, for HS generalization the gauge group is $SL(n)$. 
In the Virasoro case we define the symplectic manifold $(\mathcal{M},g)$ of \textit{all} 2d $SL(2,\mathbb{R})$ connections
\eq{
	A=A_zdz+A_{\bar{z}}d\bar{z}, 
	\label{1-ward-10}
}
with symplectic form descended from Chern-Simons (CS) action
\eq{
	\Omega = \frac{1}{2\pi\hbar}\int_\Sigma \operatorname{tr}\delta A_z\wedge \delta A_{\bar{z}}.
	\label{1-ward-11}
} 
In geometric therms $A_z$ and $A_{\bar{z}}$ are the projective connection and Beltrami differential correspondingly.
The hamiltonian reduction of the two copies of CS provides the map to the gravity theory.

The generalization from $SL(2)$ formulation to $SL(n)$ formally uses the definition of the connections as
\al{
	& A=(a^a_\mu T_a+a_\mu^{a_1\dots a_s}T_{a_1\dots a_s})dx^\mu \nonumber \\
	& \bar{A}=(\bar{a}^a_\mu T_a+\bar{a}_\mu^{a_1\dots a_s}T_{a_1\dots a_s})dx^\mu.
	\nonumber
}
Thus, the dreibeins and spin connections are defined by
\eq{
	e_\mu =\frac{1}{2}(A_\mu-\bar{A}_\mu), \qquad \omega_\mu=\frac{1}{2}(A_\mu+\bar{A}_\mu).
	\nonumber}
The action is defined in the same way as in the $SL(2)$ case, i.e. we apply all the technology developed so far to this case!
\eq{
	S_{\operatorname{grav}}=S_{CS}[A]-S_{CS}[\bar{A}] + S_{bdy}. \nonumber
}


As discussed in the previous section, we generalize as well the expression for the symplectic form of the reduced space \eqref{1-ward-11} as
\eq{
	\Omega_{red}=\frac{1}{2\pi\hbar}\int\limits_\Sigma\sum\limits_{k=2}^n\delta\mu_k\wedge \delta u_k,
	\label{red-sympl-form-1}
}
where $u_k$ is conjugate to $\mu_k$, $u_k=2\pi\hbar\delta/\delta\mu_k$.

To proceed further, we follow the arguments in \cite{hVerlinde-90,Dijkgraaf:1996iy,Li:2013rsa} and consider the inclusion of the higher spin charges. This is achieved by turning on "chemical" potentials $\mu_k$. From geometric point of view this corresponds  to deforming the symplectic form \eqref{1-ward-11} to \eqref{red-sympl-form-1}, i.e. including higher Beltrami differentials as conjugate to the conserved charges. From CFT point of view turning on the chemical potentials $\mu_k$ of the higher-spin charges can be accounted for perturbatively by adding an irrelevant perturbation to the action  
\eq{
	I_{CFT} \quad \longrightarrow\quad I_{CFT}+\int\limits_\Sigma \sum\limits_{k=2}^n \left( \mu_k Q_k - \bar{\mu}_k \bar{Q}_k \right),
	\label{irr-adds-1}
}
where $Q_s:=q_{k-3}$ and $\bar{Q}_s:=\bar{q}_{k-3}$ are the holomorphic and anti-holomorphic invariants we introduced above (and corresponding to projective rewriting of $u_k$).

As an illustration of how this procedure works, we refer to \cite{hVerlinde-90} where the $W_3$ generalization of the action has been proposed in the form
\eq{
	S_{W_3}[\mu,\lambda]=\underset{\{T,W\}}{\mathrm{min}}\left(\Gamma_{SL_3}[A(T,W)] - \int (\mu T+\lambda W) \right),
	\label{w-actions-1}
}
Here $\Gamma_{SL_3}[A(T,W)]$ is the Wess-Zumino-Witten (WZW) $SL_3$ action with reduced connection
\eq{
	A(T,W)=\begin{pmatrix}0 & 1 & 0 \\ T & 0 & 1 \\ W & 0 & 0
	\end{pmatrix}.
	\label{w-actions-1a}
}

The variation of the full action (bulk plus boundary) has been found in \cite{Li:2013rsa}  to be:
\ml{
	\delta I^{E}_{|on-shell} =\delta  I^{E}_{bulk|on-shell} +\delta  I^{E}_{bdy|on-shell} \\ =-(2\pi ik)\left(T\delta\mu -\bar{T}\delta\bar{\mu} + \sum\limits_{s=3}^N(Q_s\delta\mu_s-\bar{Q}_s\delta\bar{\mu}_s) \right).
	\label{w-actions-2}
}
Here $T$ is the stress-tensor, $\mu$ is its conjugate, $Q_s$ and $\mu_k$ are the higher charges and their conjugate (and the anti-holomorphic parts correspondingly).

The free energy is given by
\eq{
	-\beta F=S+2\pi i \left(T\mu -\bar{T}\bar{\mu}+\sum\limits_{s=3}^N (Q_s\mu_s -\bar{Q}_s\bar{\mu}_s) \right),
	\label{w-actions-3}
}
where $S$ is the entropy. A careful variation of the entropy is
\eq{
	\delta S=-(2\pi ik)\left(\delta T\mu -\delta\bar{T}\bar{\mu} + \sum\limits_{s=3}^N(\delta Q_s\mu_s-\delta \bar{Q}_s\bar{\mu}_s) \right),
	\label{w-actions-4}
}
which confirms that the entropy is a Legendre transform of the free
energy given above. Further developments along these lines can be found for instance in \cite{Perez:2016vqo}. 


Our proposal for HS generalization of the theory action is based on the construction for $SL(2)$ case, but also uses the chiral splitting of the W-gravity \cite{Dijkgraaf:1996iy,hVerlinde-90}. In a very nice paper \citen{Freidel:2008sh} Freidel has discussed the $SL(2)$ in some length. Introducing two 2d frame fields $e^\pm=e^\pm_\mu dx^\mu$ and $E^\pm$, one can show that the wave functional is given by\cite{Freidel:2008sh}
\eq{
	\Psi(e^\pm)= e^{\frac{ik}{4\pi}\int_\Sigma e} \int DE\exp\left\{-\frac{ik}{2\pi}\int_\Sigma(E^+-e^+) \wedge (E^--e^-) \right\}Z_c(E),
	\label{bulk-w-f}
} 
where, $k=\ell/4G$, $c=1+6k$, and $Z_c(E)$ satisfies 2d Ward identity. The nice output is  an explicit expression for the bulk wave functional in terms of boundary data.
The effect of integrating out the lagrangian multiplies has been explained in \cite{McGough:2016lol} and led to the action
\eq{
	S_{UV}\sim \int dudv\{x^+,u\}\{x^-,v\},
}
where the $\{x^+,u\}$ is the Schwarzian ($p_0$ in our notations) and the action is called dubbed as Schwarzian action. Note that this would correspond to a particular case of the so-called $T\bar{T}$ deformation of the original CFT. 

Following the same analogy the extension to HS looks straightforward - we have to go on the steps described by Freidel replacing the $SL(2)$ frames with $SL(n)$ ones and compute the effective action\cite{rashkov-nordita-18}.     

The systematic generalization to HS can be constructed by making use of a chain of transformation.  Let us assume that we have generating functional of correlation functions in a quadratically (double-trace) deformed theory of the form
\eq{
\mathbb{Z}[B_i,\bar{B}_i]= e^{-W[B_i,\bar{B}_i]} =\int\mathcal{D}\phi\,e^{-S[\phi]+\int B_1\bar{A}_1+\int \bar{B}_1A_1 + \int \bar B_2\bar{A}_2 +\int \bar{B}_2A_2},
\label{def-new-1}
}
where $\phi$ denotes all fundamental degrees of freedom and $B_i$ are sources with contact interaction with respect to the operators they are coupled to. It is natural to assume that in the sources are modified appropriately. The idea is to perform an inverse Hubbard-Stratonovich-like transformation\footnote{Analogous transformations has been applied, but in inverse order, in \cite{Bzowski:2018pcy}.}. To do that we introduce an auxiliary fields $(x_i,\bar{x}_i)$ with translationally invariant measure
\eq{
\operatorname{det}(a_{ij})\int\prod\limits_{i=1,2}\mathcal{D} x_i\mathcal{D}\bar{x}_i\, e^{-\int x_ia_{ij}^{-1}\bar{x}_j}=1.
\label{measure-3}
} 
Let insert the identity \eqref{measure-3} in \eqref{def-new-1} and make a shift $x_i=b_i-B_i+a_{ik}A_k, \: \bar{x}_i=\bar{b}_i- \bar{B}_i + a_{il}\bar{A_l}$
\ml{
\mathbb{Z}[B_i,\bar{B}_i]=\int\mathcal{D}\phi\,\operatorname{det}(a_{ij})\int\prod\limits_{i=1,2}\mathcal{D} x_i\mathcal{D}\bar{x}_i\,e^{-S[\phi] -\int x_ia_{ij}^{-1}\bar{x}_j +\int B_i\bar{A}_i+\int \bar{B}_iA_i}\\
= \operatorname{det}(a_{ij})\int\prod\limits_{i=1,2}\mathcal{D} b_i\mathcal{D}\bar{b}_i\,
\braket{e^{-\int \big(b_i-B_i+a_{ik}A_k\big)a_{ij}^{-1}\big(\bar{b}_j- \bar{B}_j+ a_{jl}\bar{A}_l\big) +\int B_i\bar{A}_i+\int \bar{B}_iA_i}}_0\\ 
= \operatorname{det}(a_{ij})\int\prod\limits_{i=1,2}\mathcal{D} b_i\mathcal{D}\bar{b}_i\, e^{-W[B_i,\bar{B}_i]}\,
e^{-\int \big(b_i-B_i+a_{ik}\braket{A_k}\big)a_{ij}^{-1}\big(\bar{b}_j-\bar{B}_j+ a_{jl}\braket{\bar{A}_l}\big)}\\
= \operatorname{det}(a_{ij})\int\prod\limits_{i=1,2}\mathcal{D} b_i\mathcal{D}\bar{b}_i\, e^{W[b_i,\bar{b}_i]} e^{-\int \big(b_i-B_i\big)a_{ij}^{-1}\big(\bar{b}_j-\bar{B}_j\big)},
\label{def-new-2}
}
where we introduced the generating functional
\eq{
e^{-W[b_i,\bar{b}_i]}= \int\mathcal{D}\phi\,e^{-S[\phi]+\int b_i \bar{A}_i + A_i\bar{b}_i + A_ia_{ij}\bar{A}_j}.
\label{def-f-1}
} 

Solving the integral by using the saddle-point approximation  gives
\al{
& \frac{\delta W[b_i,\bar{b}_i]}{\delta b_i}-a_{ij}^{-1}(\bar{b}_j-\bar{B}_J)=0\quad \Rightarrow \quad \bar{b}_i=a_{ij}\braket{\bar{A}_j}+\bar{B}_j, \label{saddle-1a} \\
& \frac{\delta W[b_i,\bar{b}_i]}{\delta \bar{b}_i}-a_{ij}^{-1}(b_j-B_J)=0\quad \Rightarrow \quad b_i=a_{ij}\braket{A_j}+B_j,
\label{saddle-1b}
}
where we used that $\delta W[b_k,\bar{b}_k]/\delta b_i=\braket{\bar{A}_i}$ and $\delta W[b_k,\bar{b}_k]/\delta \bar{b}_i=\braket{A_i}$. 

If we consider $a_{22}\neq 0$ and all other vanishing, and $A_2=T$, $\bar{A}=T$ we find the so-called $T\bar{T}$ integrable deformation. In the case Virasoro symmetry this is double-Schwarzian deformation, i.e. $SL_2$ projective invariant. The deformation to the initial Lagrangian has the form (in our notations)
\eq{
\delta \mathcal{L}=a_{22}\,q_0\bar{q}_0.
\label{def-vir-1}
}
Our proposal for "$T\bar{T}$" deformation of $SL_3$ HS theory, i.e. W-symmetry, is
\eq{
\delta \mathcal{L}=a_{33}\,q_1\bar{q}_1.
\label{def-W-1}
}
The expression for the $T\bar{T}$ is modified correspondingly since we have higher spin theory.

Note that the choice of $a_{12}\neq 0$ and all others vanishing leads to a deformation term of the form $J\bar{T}$ (or the hermitean conjugate for $a_{21}\neq 0$). 


In the case of higher projective invariants the simplest case is when only $q_{n-2}$ and $\bar{q}_{n-2}$ are turned on. This means that we have chosen $a_{nn}\neq0$ and all other $a_{ij}$ vanishing. 
Following  the discussion above our proposal for $T\bar{T}$ deformation of the HS Lagrangian in this case is
\eq{
\delta\mathcal{L}= a_{nn}\,q_{n-2}\bar{q}_{n-2}.
\label{hs-gen-1}
}
According to expressions in Section \ref{sect-3} the invariant $q_{n-2}$ is given by
\eq{
q_{n-2}=\frac{n-1}{2}\left(\frac{W_n'}{W_n}\right)'- \frac{(n-1)}{2n} \left(\frac{W_n'}{W_n}\right)^2 +\frac{W_{n-2}}{W_n}. 
}
and reduced symplectic form in these notations becomes
\eq{
\Omega_{red}=\frac{1}{2\pi\hbar}\int\limits_\Sigma\sum\limits_{k=2}^n\delta\mu_k\wedge \delta q_{k-2}\,\equiv \frac{1}{2\pi\hbar}\int\limits_\Sigma \delta\mu_n\wedge \delta q_{n-2}.
\label{red-sympl-form-2}
}

We interpret equation \eqref{hs-gen-1} as "double Schwarzian" for higher spin theories. 
Note that for the case $n=2$ the expression \eqref{hs-gen-1} reduces to the double Schwarzian theory discussed in \cite{McGough:2016lol}. 


\section{Comments on the relations to integrable systems}\label{sect-5}

In this Section we present more direct links to integrable systems, namely relations between invariants and tau-functions.  The most relevant language for this purpose is that of KP hierarchies. 

First of all, let us recall the generic form of the  KP operator
\eq{ 
\mathcal{D}_k=\left(\pa^n+u_{n-1}\pa^{n-1}+\dots+u_0 \right),
\label{KP-oper}
}
which has the same form as the that in \eqref{ode-w-12}. It has deep meaning which can be unraveled looking at the associated geometric picture, namely Sato grassmannian \cite{sato-grassmann}. The simplest  differential operator $\pa^n$ (i.e. $u_i=0$ for all $i$) corresponds to choosing the trivial point of the Sato grassmannian, 
\eq{
	\mathcal{W}_0=\{1,\lambda,\lambda^2,\lambda^3,\dots \}.
}
But the operator $\mathcal{D}_k$ \eqref{KP-oper}, being in a general form corresponds to  a generic point of the grassmannian. Thus, it should be possible to lift a function on the
Riemann surface to a section of some bundle over the Riemann surface with the
fiber equivalent to the grassmannian.
To make all these explicit, let us start mentioning the Fourier-Laplace transform of a function $f(x)$
\eq{
	f(x)=\int e^{x\lambda}\hat{f}(\lambda) d\lambda.
}
Next step is to lift this function to a \textit{section of a certain bundle over the grassmannian}. As it is well known, there exist some sections, which are defined by KP
flows
\eq{
	f(t_1,\dots,t_n)=\int e^{t_1\lambda\,+ \dots +t_n\lambda^n\, + \dots} \hat{f}(\lambda)d\lambda,
	\label{section-f}
}
with $t_1=x$. Obviously
\eq{
	\pa^n_x f\equiv\pa^n_{t_1}f=\pa_{t_n}f.
	\label{section-f-1}
}

This however, is when derivative acts as KP flow at the trivial point $\mathcal{W}_0$. To move to a generic point one should use the transform
\eq{
	f(t_1,\dots,t_n)=\int \Psi_{\mathcal{W}}(\lambda,\{t_n\}) \hat{f}(\lambda)d\lambda,
	\label{section-f-2}
}
where $\Psi_{\mathcal{W}}(\lambda,\{t_n\})$  is some Baker-Akhiezer function.

Thus, \eqref{section-f-1} takes the general form
\begin{equation}
\pa_{t_n}f=\left(\pa^n+u_{n-1}\pa^{n-1}+\dots+u_0 \right)f.
\label{KP-flow-1}
\end{equation}
The consistent truncation to the n-th flow, $\pa_{t_n}f=0$ leads to the equation
\eq{
\left(\pa^n+w_1\pa^{n-1}+\dots+w_n\right)f=0. 
\label{eq-tau-1}
}
From one hand side the Baker-Akhiezer function associated with the hierarchy is given by
\al{
	\Psi_{\mathcal{W}}(\lambda,\{t_n\}) & = e^{\sum_nt_n\lambda^n}\frac{\uptau(t_1-1/\lambda, \dots, t_n-1/(n\lambda^n),\dots )}{\uptau(t_1,\dots,t_n,\dots)} \nonumber\\
	& = e^{\sum_nt_n\lambda^n}\cdot \left[1+\sum\limits_{i=1}^\infty w_i(\{t_n\})\lambda^{-i} \right],
	\label{KP-BA}
}
where $\uptau(x;\{t_n\})$ is the celebrated tau-function. On the other hand, the tau-function associated with the solutions of the equation \eqref{eq-tau-1} is the determinant
\eq{
\uptau_n(x)=\operatorname{det}|\pa^{i-1}f_k|=\operatorname{det}\begin{vmatrix}
f_1 & \cdots & f_n\\f_1'& \cdots & f_n'\\\vdots & & \vdots\\f_1^{(n-1)} & \cdot & f_n^{(n-1)}
\end{vmatrix}, 
\label{eq-tau-2}
}
where $f_i$ are the solutions of \eqref{eq-tau-1}. These relations provide direct link between higher projective invariants and the corresponding  tau-function.

To obtain the relation between $w_k$ in \eqref{KP-BA} and $u_k$ in \eqref{KP-oper} one has to take logarithm of \eqref{KP-BA}, expand both sides in series of $1/\lambda$ and use KP flow equations (residue w.r.t. $\pa$, i.e. the coefficient of $\pa^{-1}$)
\eq{
	\pa_{t_1}\pa_{t_n}\log\uptau=\left(L^n\right)_{-1},
	\label{KP-flow-2}
}
where $L$ is
\eq{
	L = \pa + a_2 \pa^{-1} + a_3 \pa^{-2} + a_4 \pa^{-3} + \cdots .                 
	\label{4.2}
}  
As an example, for the first two reductions one finds
\al{
	& SL(2)\:\:\text{reduction}\:\: \pa_{t_2}\log\tau=0:\qquad w_2=u_2, \\ 
	& SL(3)\:\:\text{reduction}\:\: \pa_{t_3}\log\tau=0:\qquad w_3=\frac{1}{2}u_2'+u_3.
\label{KP-reduction}
}

To this end it is useful to introduce the Hirota notations for tau-functions and the invariants $w_j$\footnote{In our notations the numbers in the formulas below stand for the order of derivatives in the corresponding row.}
\eq{
	\uptau(x;t)=|0,1,2,\dots,m-1|,
	\label{7.2}
}
and
\eq{
	w_j(x;t)=(-1)^j\frac{|0,1,\dots,m-j-1,m-j+1,\dots,m|}{|0,1,\dots,m-1|}.
	\label{7.3}
}
To take advantage of these notations we consider the partition $[\underset{m}{\underbrace{1,1,\dots,1}},\underset{m-j}{\underbrace{0,0,\dots,0}}]$, or the Young diagram with $j$ boxes in one column $\Yboxdim{4pt} \yng(1,1,1,1)$. Then from the definition of the tau-function it follows that
\eq{
	S_{\Yboxdim{4pt} \yng(1,1,1,1)}(\tilde{\pa}_t)\tau=
	|0,1,\dots,m-j-1,m-j+1,\dots,m|,
	\label{7.12}
}
where  $S_{\Yboxdim{4pt} \yng(1,1,1,1)}(\tilde{\pa}_t)$ \footnote{Here $\tilde\pa_t$ denotes
 $\tilde\pa_t:=\left(\frac{\pa}{\pa t_1}, \frac 12\frac{\pa}{\pa t_2}, \frac 13\frac{\pa}{\pa t_3}, \cdots
\right)$.} is the Schur function. 
After plugin the above result into \eqref{7.3} we find
\eq{
	w_j=(-1)^j\frac{1}{\uptau} S_{\Yboxdim{4pt} \yng(1,1,1,1)}(\tilde{\pa}_t)\uptau.
	\label{7.13}
}

This expression provides an explicit relation between higher invariants and tau-functions.

At this point we would like to make a few comments. First of all we remind that the action for 3d gravity with negative cosmological consists of the difference of.  two Chern-Simons actions. For each of them one can define higher projective invariants. Having in mind the arguments in Section \ref{sect-2} we conjecture that the tau-functions involved the above expressions are actually Toda tau-functions. Indeed, they looks pretty much like double-wronskian solutions.


\section{Concluding remarks} 

It is strongly believed that integrability plays important role in holographic correspondence. Particular structures  of integrable hierarchies appear in many phenomena, especially those associated with strong interactions. Many aspects of string theory side of the correspondence are naturally related to integrable system, but it is conjectured that the later are also building blocks of HS theories involved in low-dimensional holography.
In this paper we considered several aspects of low-dimensional holography from point of view of integrable structures. 

As motivating example of such structures we consider entanglement entropy of a single interval. It turns ut that the expression for the entanglement entropy can be considered from three different angles. We found three expressions for it. 

The first one gives an expansion in terms of Aharonov invariants. These are invariant with respect to M\"obius transformations and explicit representation in  terms of the Schwarzian and its derivatives. 

The second expression involves Faber polynomials, which are defined by choosing appropriate basis in the space of univalent functions. As we discussed in Section \ref{sect-2}, the excited states are realized by an unitary operator $U_f$ corresponding to conformal transformation $z\to w=f(z)$. An important structures here are also Grunsky coefficients, which provided a link to the next expression.

The third expression we found realizing that Faber polynomials can be mapped to the symbols of certain Lax operator of dispersionless Toda integrable hierarchy.  Analyzing the expressions we found that the Grunsky coefficients are in fact second derivatives of Toda tau-function.

The conclusions from Section \ref{sect-2} are that entanglement entropy is related to the $SL_2$ projective invariant (Schwarzian) and its derivatives. On the other hand, it is related to dToda hierarchy and its tau-functions.

In Section \ref{sect-3} we present explicit expressions for $SL_n$ projective invariants (higher Schwarzians) with the idea to generalize the picture to HS theories. The final expressions are nicely organized into (sum of) partial Bell polynomials. The comparison of our expressions for the case of $SL_3$ agrees exactly with know results and we provided a short comments on the relation to old works on the origin of W-algebras.

The next Section \ref{sect-4} considers integrable deformations of HS action by invariant operators. The deformation with a single projective invariant is suggested to be interpreted as higher Schwarzian theory (or HS Schwarzian theory). In remarkable works  Zamolodchikov \cite{Zamolodchikov:2004ce} and Smirnov and Zamolodchikov \cite{Smirnov:2016lqw} considered deformations by irrelevant operators, which attracted quite some attention. It has bee shown that these deformations have important physical consequences. For instance, it has been demonstrated that deformed and undeformed theories are related by field-dependent coordinate transformations \cite{Conti:2018tca}. On the other hand, one can show that they lead to Ban\~ados-type metrics. In another paper different deformations with broken Lorentz invariance (bur still integrable) has been suggested \cite{Bzowski:2018pcy}. In this Section we suggested expressions for such deformations in the case of HS theories.

In Section \ref{sect-5} we provide comments on invariants we have found from point of view of integrable hierarchies. First, one can easily see that the space of univalent functions considered in Section \ref{sect-2} can be mapped to the trivial point of the universal Sato grassmannian. From this perspective the appearance of tau-functions looks natural. The determinant representation of the projective invariants also points toward representation in terms of tau-functions. In this Section we give explicit relation between these quantities.  

The development of findings in this study can be pursued in many directions. Although quite some relations have been unraveled in this paper, many physical consequences and questions remain to be addressed. First of all, it would be interesting to consider the boundary conditions on the HS gravity side for irrelevant deformations. The role of such deformed theories as UV completion certain QFT's just began. This knowledge could provide better understanding of the HS holography. Another interesting issue is the geometric meaning behind such deformations in higher spin theories and W-geometries in particular. Indeed, applications to the minimal surfaces in the context of holographic correspondence could be important. As it has already  been  pointed for instance in \cite{Perez:2016vqo}, the KdV hierarchy plays important role in low-dimensional holography.  On the other hand, tau-functions are powerful tool in studying main features on both sides of correspondence – correlation functions, partition functions etc. Thus, it remains to thoroughly develop these ideas along the lines above. Another direction is the study of W-minimal surfaces by making use of tau-functions. Although flag structure in our context is more or less well studied, the physical consequences of the deformations of flags still needs investigation. Finally, the role of the information geometry and quantum information for bulk reconstruction is sharply increasing. It is not quite clear whether (and how) the information geometry account for higher structures in this context.
We plan to address some of these issues in future publications.

\paragraph{Acknowledgments}\ \\

 I thank Hamid Afshar, H.~Dimov and especially Daniel Grumiller for comments and critically reading the manuscript. I'm grateful to Daniel Gruniller for bringing to my attention \cite{Ecker:2019ocp}. 
This work was finalized during the program “Higher Spins and Holography 2019” at the Erwin Schrodinger International Institute (ESI). I would like to thank ESI and organizers for warm hospitality and stimulating atmosphere. This project was supported in part by BNSF Grant DN-18/1 and H-28/5.

\begin{appendix}
\section{Bell's polynomials and some notations}

We used Bell's polynomials in the text to represent invariants in convenient form. The generating function of Bell's polynomials is
\ml{
	\Phi(t,x)\equiv \exp\left(u\sum\limits_{m\geq 1}x_m\frac{t^m}{m!} \right)= \sum\limits_{n,k\geq 0}B_{n,k}(x_1,x_2\dots)\frac{t^n}{n!}u^k \\
	=1+ \sum\limits_{n\geq 1}\frac{t^n}{n!}\left\{\sum\limits_{k=1}^n u^k B_{n,k}(x_1,x_2\dots) \right\}.
	\label{generating-bell}
}

For $n$ and $k$ non-negative integers, the (exponential) $(n,k)$ partial Bell polynomial in the variables $x_1,x_2,\dots, x_{n-k+1}$ are denoted by $B_{n,k}\equiv B_{n,k} (x_1, x_2,\dots, x_{n-k+1})$. They may be defined by the formal power series expansion
\eq{
	\frac{1}{k!}\left(\sum\limits_{m=1}^\infty x_m\frac{t^m}{m!} \right)^k=\sum\limits_{k=n}^\infty B_{n,k}(x_1, x_2,\dots, x_{n-k+1})\frac{t^n}{n!}, \qquad k\geq  0.
	\label{bell-def}
}

The partial or incomplete exponential Bell polynomials are a triangular array of polynomials. Their the explicit form is
$$
{\displaystyle B_{n,k}(x_{1},x_{2},\dots ,x_{n-k+1})=\sum {n! \over j_{1}!j_{2}!\cdots j_{n-k+1}!}\left({x_{1} \over 1!}\right)^{j_{1}}\left({x_{2} \over 2!}\right)^{j_{2}}\cdots \left({x_{n-k+1} \over (n-k+1)!}\right)^{j_{n-k+1}},}
$$
where the sum is taken over all sequences $j_1,j_2,\dots, j_{n-k+1}$
of non-negative integers provided the following two conditions are satisfied:
\al{ 
	& {\displaystyle j_{1}+j_{2}+\cdots +j_{n-k+1}=k,}
	\\
	& {\displaystyle j_{1}+2j_{2}+3j_{3}+\cdots +(n-k+1)j_{n-k+1}=n.}
}

Below we list Bell's polynomials $B_{n,k}$ for the first several $n$:
\begin{itemize}
	\item List of $B_{3,k}(x_1,x_2,x_3)$
	\eq{
		B_{3,1}(x_1,x_2,x_3)=x_3, \quad B_{3,2}(x_1,x_2,x_3)=3x_1x_2, \quad B_{3,3}(x_1,x_2,x_3)=x_1^3
	}
	\item  List of $B_{4,k}(x_1,x_2,x_3,x_4)$
	\al{
		& B_{4,1}(x_1,x_2,x_3,x_4)=x_4, \qquad B_{4,2}(x_1,x_2,x_3,x_4)=3x_2^2 +4x_1x_3 \nonumber \\
		& B_{4,3}(x_1,x_2,x_3,x_4)=6x_1^2x_2, \qquad B_{4,4}(x_1,x_2,x_3,x_4)=x_1^4
	}
	\item  List of $B_{5,k}(x_1,x_2,x_3,x_4,x_5)$
	\al{
		& B_{5,1}=x_5, \quad B_{5,2}=10x_2x_3 +5x_1x_4 \nonumber \\
		& B_{5,3}=15x_1x_2^2+10x_1^2x_3, \nonumber \\ & B_{5,4}=10x_1^3x_2, \quad B_{5,5}=x_1^5
	}
	\item List of $B_{6,k}(x_1,x_2,x_3,x_4)$
	\al{
		& B_{6,1}=x_6, \qquad B_{6,2}=10x_3^2+15x_2x_4+6x_1x_5,\nonumber \\ 
		& B_{6,3}= 15x_3^2+60x_1x_2x_3 +15x_1^2x_4 \nonumber \\
		& B_{6,4}=45x_2x_2^2+ 20x_1^3x_3, \nonumber \\  & B_{6,5}=15x_1^4x_2, \qquad B_{5,5}=x_1^6
	}
\end{itemize}

In the text we used the following definition of the Pochhammer symbol:
$$
{\displaystyle (x)_{n}=x^{\underline {n}}=x(x-1)(x-2)\cdots (x-n+1)=\prod _{k=1}^{n}(x-(k-1))=\prod _{k=0}^{n-1}(x-k).}
$$

\end{appendix}


\end{document}